\documentclass[prb,aps,preprint,epsfig]{revtex4}
\usepackage{graphicx}
\usepackage{dcolumn}
\usepackage{bm}
\begin{document}
\title {Resistivity noise in crystalline magnetic nanowires and its
implications to domain formation and kinetics}
\author{Amrita Singh\footnote[1]{electronic mail:amrita@physics.iisc.ernet.in}}
\author{Debtosh Chowdhary}
\author{Arindam Ghosh}
\address{Department of Physics, Indian Institute of Science, Bangalore 560 012, India}
\begin{abstract}
We have investigated the time-dependent fluctuations in electrical resistance, or noise, in high-quality crystalline magnetic nanowires within
nanoporous templates. The noise increases exponentially with increasing temperature and magnetic field, and has been analyzed in terms of domain
wall depinning within the Neel-Brown framework. The frequency-dependence of noise also indicates a crossover from nondiffusive kinetics to
long-range diffusion at higher temperatures, as well as a strong collective depinning, which need to be considered when implementing these
nanowires in magnetoelectronic devices.
\end{abstract}


\maketitle

In recent years, magnetic nanostructures have attracted much interest due to their rich physics and potential application in high density
recording and storage,~\cite{Nielsch} memory,~\cite{Parkin_Science_2008} or logic devices.~\cite{Allwood} Crystalline magnetic nanowires in
self-organized nanoporous templates are of particular interest in this context, since they are not only attractive as perpendicular storage
medium,~\cite{Nielsch} but may also form the backbone to various designs of nonvolatile domain wall (DW)-based dynamic memory. The latter relies
on the spin torque effect, where a spin-polarized current can manipulate both the location and movements of the DW, and has been widely
implemented in magnetic multilayers,~\cite{Buhrman} or in patterned nanowires from magnetic thin
films.~\cite{Allwood,Grollier,Moriya,yamaguchi_PRL_2004,Hayashi,Beach,Atkinson,Laufenberg} However, in spite of potential application in 3D
micoelectronics,~\cite{Parkin_Science_2008} investigations of DW kinetics in templated nanowires are limited, and many experiments have assumed
these systems to be single domain nanomagnets below a critical diameter of $\sim 55$~nm.~\cite{Nielsch}



Here we have employed measurement of random time-dependent fluctuations, or
noise, in electrical resistance of Nickel (Ni) nanowires to investigate both
formation as well as kinetics of the DWs as function of temperature ($T$) and
external magnetic field ($H$) applied parallel to the nanowire axis.
Experiments with spin glass,~\cite{Israeloff} magnetic sensors,~\cite{Jiang}
tunnel junctions,~\cite{Pannetiera} and ferromagnetic thin
films~\cite{Giordano,Ravelosona_PRL_2005} have already established the dominant
role of spin fluctuations and DW relaxation on electrical noise, where the DWs
have been been viewed as scattering centers of spin-polarized electrons. In
case of nanowires such measurements do not exist in a systematic manner, and
may provide an independent tool to study the magnetization reversal and domain
propagation in presence of pinning and disorder in magnetic nanostructures.


Nickel nanowires (NiNWs) were electrodeposited in nanoporous polycarbonate or
anodic alumina templates with average interpore distance $\sim 130$~nm. The
diameter of the nanowires were $\approx 50$~nm ($\pm 10\%$), below which the
Ni-nanowires have been assumed single-domain.~\cite{Nielsch} The high aspect
ratio of $120 - 1000$ in our NiNWs ensures magnetization along the wire axis
due to shape anisotropy. The cross-sectional scanning electron micrograph of
nanowires, embedded inside an alumina template, is shown in Fig.~1a. Fig.~1b
displays the XRD pattern (Philips X"PERT Pro Diffractometer) of an ensemble of
NiNWs, which confirms that the nanowires are single crystalline and stabilized
in face centered cubic (fcc) phase, with growth direction [111]. The
accompanying traces of nickel-oxide were caused during partial etching of the
template with phosphoric acid prior to the XRD measurement. Magnetic
characterization of NiNWs was done by vibrating sample magnetometer (VSM 7300,
Lakeshore) with magnetic field applied parallel to the nanowire axis. In
Fig.~1c, the $M-H$ hysteresis loop of NiNWs shows that the coercive field
$(H_{c} = 100$~Oe) and the retentivity $(M_{r} = 20 \%)$ are both much smaller
than that from Stoner-Wohlfarth model. This can arise from a strong dipolar
interaction due to the close, often inhomogeneous, packing of the nanowires
within the templates, as well as finite sample dimensions.~\cite{Rahmana}

For electrical measurements, the NiNWs were grown after coating the membranes
on one side with thin gold film, which also acts as cathode during
electrodeposition (inset of Fig.~2a). The time-dependent fluctuations in the
nanowire resistance were measured in a dynamically balanced Wheatstone bridge
configuration, followed by amplification, digitization and digital signal
processing.~\cite{Singh,Chandni} The $T$ dependence of resistance {$R$} of the
NiNWs shows metallic behavior down to $T = 4.2$~K (not shown) which, following
Ref.[18], also allowed evaluation of approximate number of nanowires between
the metallic leads ($\approx 1100$ in the device presented here). For all noise
measurements, $T$ was held constant with an accuracy of $\sim 1$~ppm, and the
measurement current density was chosen to be ${1.4\times10^{4}}$ {A/cm$^{2}$}.
Fig.~1d shows typical time series of resistance fluctuations (after subtracting
the mean) for two temperatures ($T = 80$~K and 300~K) at $H = 0$~T, which
clearly demonstrate the influence of thermal fluctuations.

%

Fig.~2 shows the evolution of the power spectral density (PSD), $S_R$, of
resistance noise as a function of $T$ at $H = 0$~T for a set of pristine
freshly grown nanowires in polycarbonate templates. First, the functional form
of $S_R \sim 1/f^\alpha$, indicates a wide distribution of time scales in the
relaxation mechanism that produce noise.~\cite{Eberhard} Moreover, an
exponential increase in the noise magnitude (at all frequency $f$) with
increasing $T$ indicates such time scales follow activated kinetics (Fig.~4a).
An immediate explanation of this observation could be found in the
pinning-depinning of magnetic DWs, where the pinning lifetime $\tau$ determines
the overall ``diffusivity'' of the DWs, and hence the magnitude of noise.
Within the Neel-Brown framework of magnetization reversal,

\begin{equation}
\label{eq1} \tau^{-1} = f_0\exp{\left[-E(H)/k_BT\right]}
\end{equation}

\noindent where $f_0$ is the attempt frequency, and $E(H) = E_0 - 2M_sVH$ is the field dependent energy barrier ($E_0$, $M_s$ and $V$ are the
bare pinning potential, saturation magnetization and activation volumes, respectively). There have been several experimental reports on
Neel-Brown relaxation in DW depinning in particular for thin magnetic films,~\cite{Attane} which suggest the DWs to be pinned at microtwins with
a wide distribution of depinning lifetimes arising from a broad density of states in $E_0$. Defects in the form of microtwins and dislocations
have already been shown to form in electrodeposited nonmagnetic nanowires as well.~\cite{Singh}


In order to further probe the validity of Neel-Brown scenario in our case, we have subsequently measured the resistance fluctuations as a
function of $H$ at $T = 300$~K. As shown in Fig.~3, increasing $H$ from 0 to $\pm 2000$~Gauss leads to a nearly three orders of magnitude
enhancement in noise magnitude (expressed as normalized variance $\delta R^2/R^2 (= \int df S_R(f)/R^2$). Although the behavior of noise is
symmetric about $H = 0$, the hysteresis between the forward and backward sweeps indicate a significant irreversible component in the
magnetization reversal process.~\cite{Chandni} However, the overall behavior of noise can be readily understood from a field-induced increase in
$\tau^{-1}$, and also provides a reasonable estimate of the activation volume $V$ from $S_R \sim \sqrt{D_{DW}} \sim \tau^{-1/2}$, where $D_{DW}$
is the DW diffusivity.~\cite{Fourcade} Taking a factor of $\approx 500$ increase in noise at $H = 2000$~Gauss, and $M_s \approx 480$~emu/cc, we
find $V \approx (6~\mbox {nm})^3$, indicating nucleation to take place around a localized defect, rather than the entire nanowire
volume.~\cite{Attane}

%

With both $T$ and $H$ dependence of noise strongly indicating existence of mobile DWs, we now focus on two intriguing features of the zero-field
noise that reveals the stochastic nature of DW propagation in our NiNWs. First, we note from Fig.~2 that the frequency exponent $\alpha$ of PSD
is $T$-dependent itself. The variation of $\alpha$ which, as shown in Fig.~4b, gradually changes from $\sim 1$ to $3/2$ as $T$ is increased. It
has been shown both theoretically~\cite{Fourcade} and experimentally~\cite{Koch} that the latter limit is a direct manifestation of long-range
diffusion of the scattering species through a percolative or fractal network. Indeed, such a scenario can be envisaged for magnetic systems as
well arising from an interplay between magnetic anisotropy and demagnetization fields.~\cite{Attane,Sayko} In our case, the transition of
$\alpha$ from $1$ to $3/2$ thus indicates a crossover from hopping to diffusive kinetics of the DWs in magnetic nanowires with increasing $T$.
The influence of measuring current, which is kept several orders of magnitude lower than its typical critical magnitude ($\sim 10^8$~A/cm$^2$)
for DW motion is unclear at present, and a measurement of noise as a function of both excitation current and temperature is in progress.

Second, the noise data also probes the collectivity of scatterer (DWs) kinetics
by investigating the Gaussianity of noise. Fig.~4c shows a statistics of the
resistance jumps ($\delta R$) between two successive states,~\cite{Chandni} for
four different values of $T$ (at $H = 0$~T). At low $T$, where DW kinetics is
nondiffusive, the noise statistics is essentially Gaussian, but as $T$ is
increased above $\sim 260$~K a non-Gaussian tail becomes evident. Such a
non-Gaussianity in noise indicates the DW movements inside the nanowire are
highly correlated, and expectedly becomes significant when the DWs undergo
long-range diffusion. This may be of significance in DW-based device
applications, where the manipulation of a particular domain may have a
considerable influence on the magnetization and kinetics of other domains in
proximity as well.

In summary, we have probed the formation and kinetics of domain walls in
electrodeposited Nickel nanowires by electrical noise measurement. The
temperature and magnetic field dependence of noise measurement are consistent
with thermally assisted depinning of domain walls, and can be analyzed within
the Neel-Brown framework. We also found the domain wall kinetics to be
diffusive and strongly correlated at higher temperatures.

\textbf{Acknowledgement} We thank Ministry of Communication and Information
technology, Government of India, for a funded project and Prof. A. Sundaresan,
Jawaharlal Nehru Centre for Advanced Scientific Research, for the VSM
measurements.

\clearpage

\begin{figure}[t]
\vspace{-.2cm}\caption{Color Online. (a) Cross-sectional scanning electron
micrograph, (b) X-Ray diffraction pattern, (c) $M-H$ hysteresis loop (with
magnetic field applied parallel to the axis of nanowires, as shown in the
inset), of Ni nanowires grown inside nanoporous templates. (d) Time series of
resistance fluctuation, at two different temperatures.} \label{figure1}
\end{figure}

\begin{figure}[t]
\vspace{-.2cm}\caption{Color Online. Normalized power spectral density of noise
at different temperatures. The spectral exponent ($\alpha$) decreases with
decrease in temperature from $300$ K to $80$ K. The inset shows the schematic
of contacted nickel nanowires.} \label{figure2}
\end{figure}

\begin{figure}[t]
\vspace{-.2cm}\caption{Color Online. The evolution and hysteresis of noise
magnitude with magnetic field at $300$~K. The field is applied parallel to
nanowires axis. The arrows indicate direction of field change.} \label{figure3}
\end{figure}

\begin{figure}[t]
\vspace{-.2cm}\caption{Color Online. (a) Normalized variance of noise as a
function of temperature. Inset: Same in thermally activated form, which gives a
bare pinning potential of $\approx 168$~mV in the present case. (b). Variation
of spectral exponent ($\alpha$) with temperature. (c) Variation of probability
distribution function  of resistance jumps in nickel nanowires at four
different temperatures, which shows that the statistics becomes non-gaussian at
higher ($\sim 300$ K) temperatures.} \label{figure4}
\end{figure}


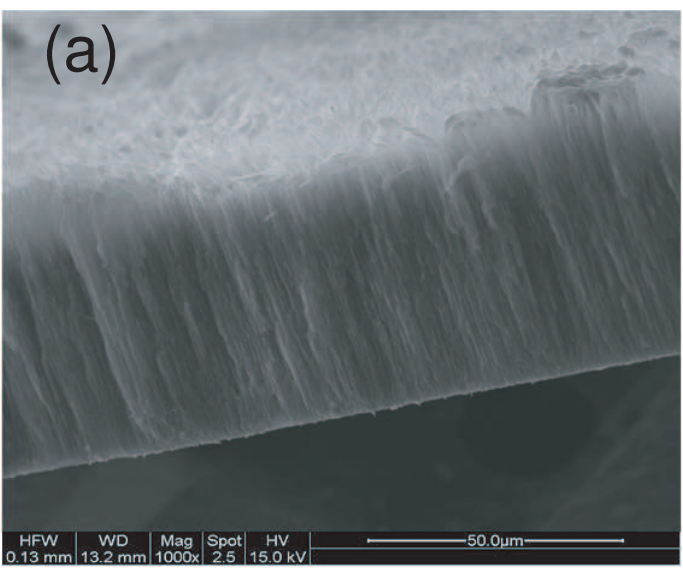
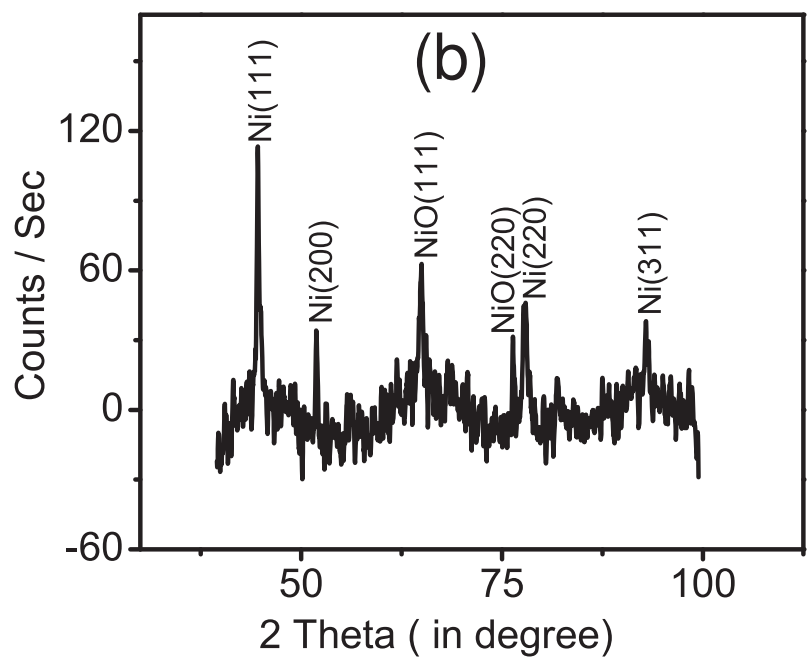
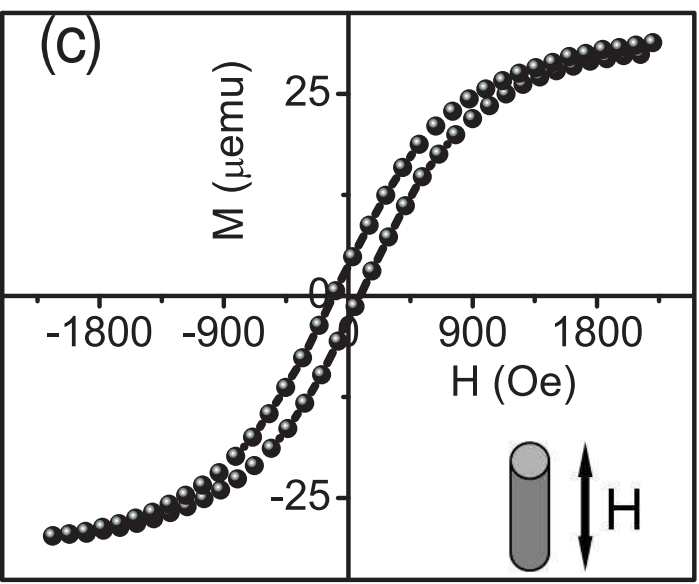
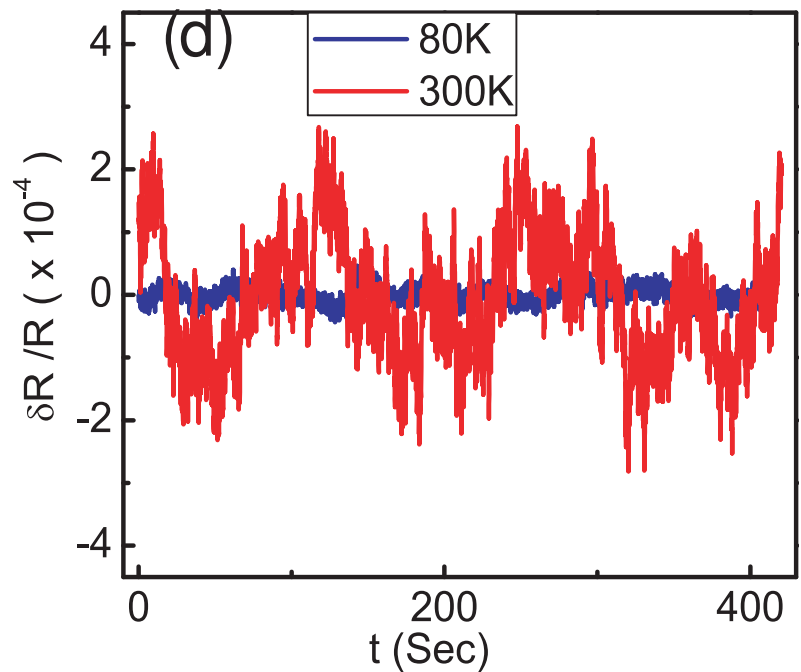
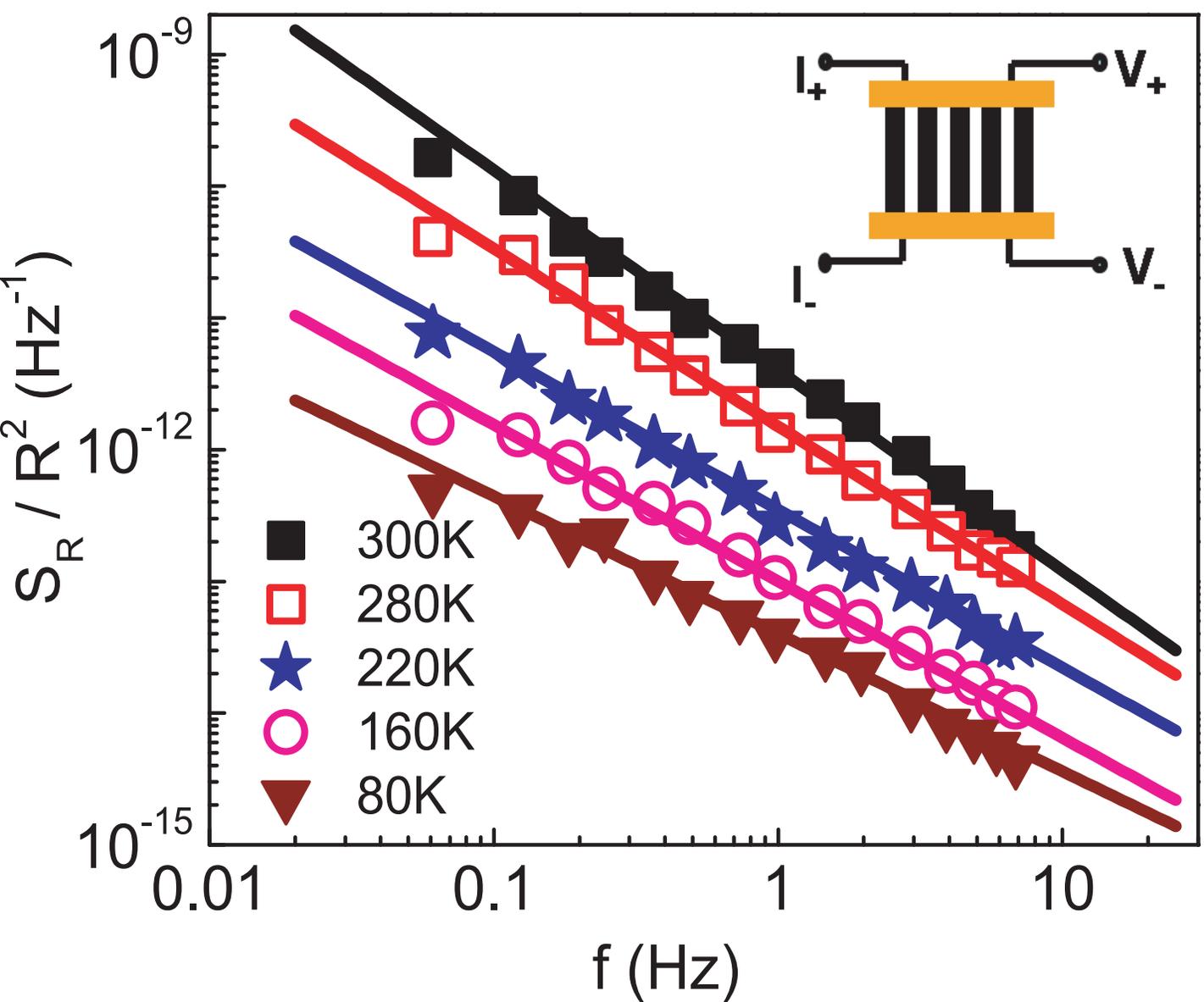
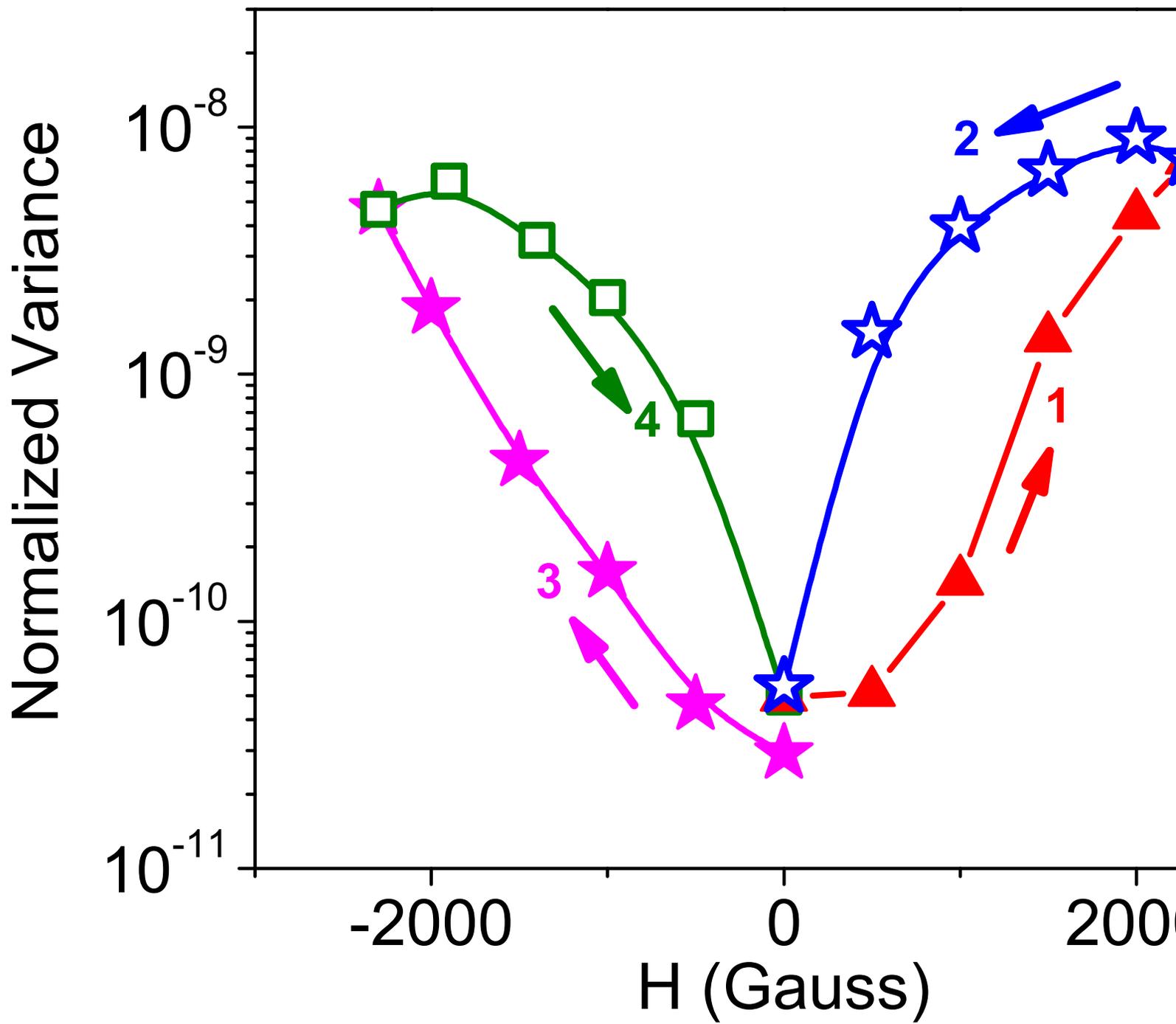
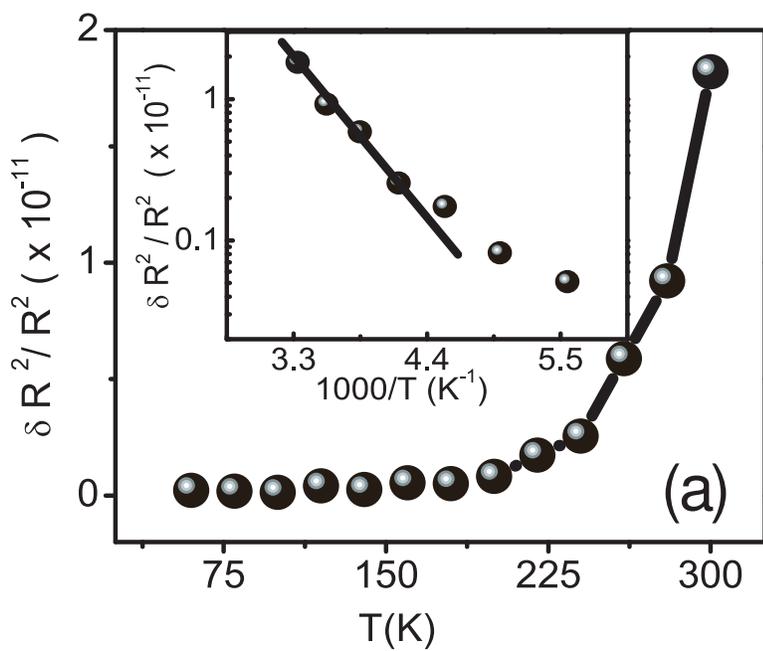
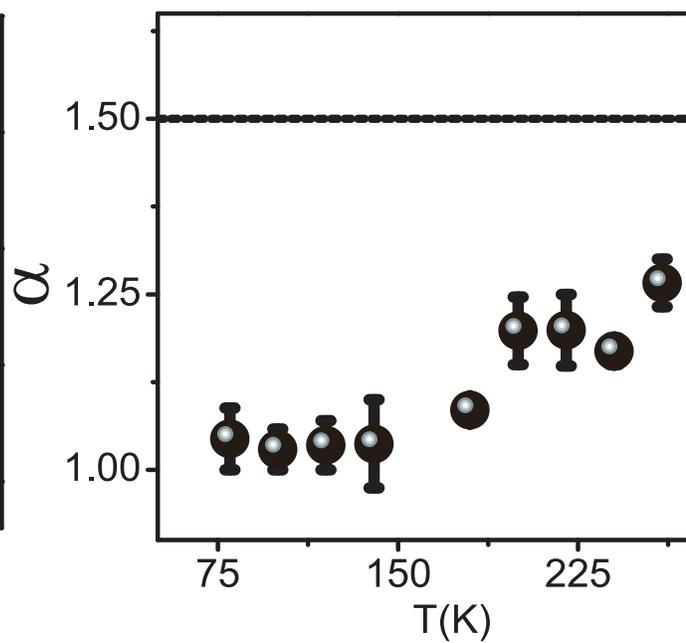
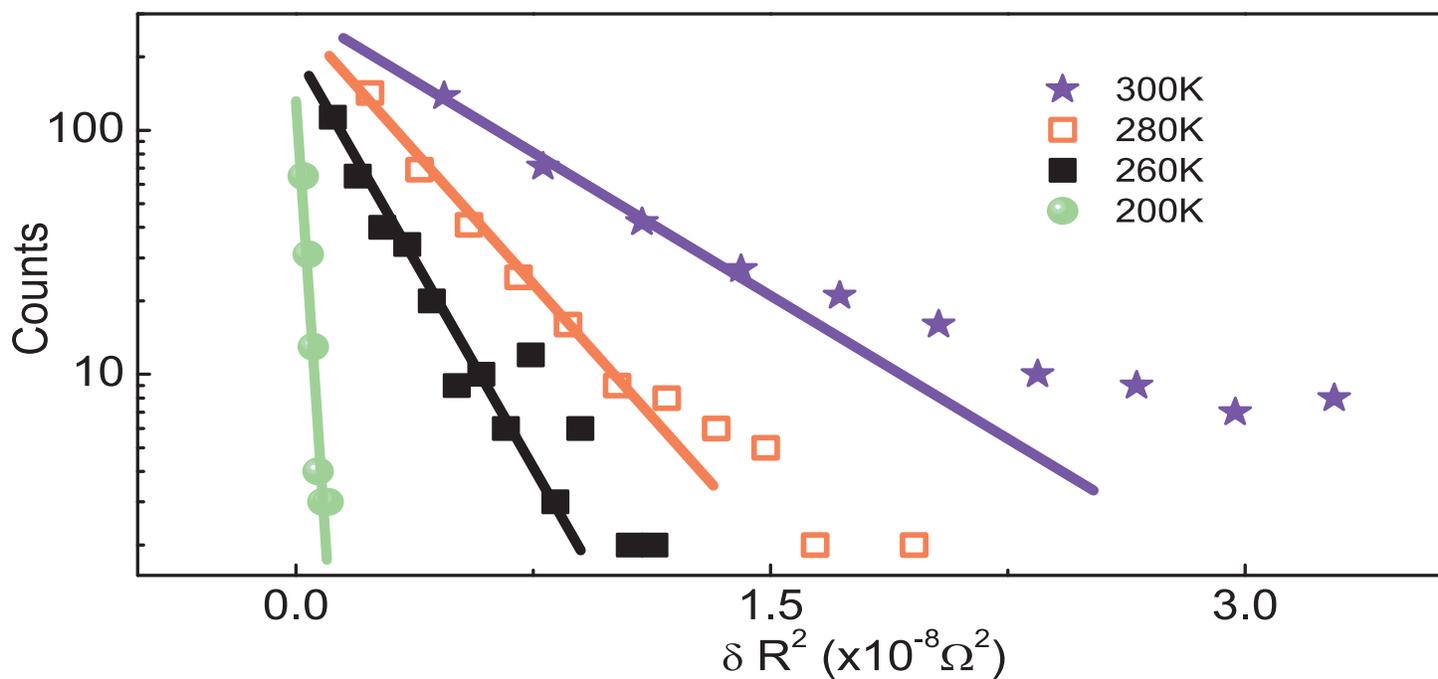
\end{document}